# Strong Temperature Dependence of Thermal Conductivity in High-Purity Cubic Boron Arsenide

Songrui Hou[1], Fengjiao Pan[2], Xinping Shi[3], Frank Angeles[3], Geethal Amila Gamage[2], Haoran Sun[2], Benise A. Niyikiza[2], Zahra Ebrahim Nataj[4], Fariborz Kargar[4], Alexander A. Balandin[4], David G. Cahill[5], Chen Li[1,3]*, Zhifeng Ren[2]*, Richard B. Wilson[1,3]*

**Affiliations:**

[1]Materials Science and Engineering, University of California, Riverside, CA, 92521, USA.

[2]Department of Physics and Texas Center for Superconductivity (TcSUH), University of Houston, Houston, TX 77204, USA.

[3]Department of Mechanical Engineering, University of California, Riverside, CA, 92521, USA.

[4]Department of Materials Science and Engineering, University of California, Los Angeles, California 90095, USA.

[5]Department of Materials Science and Engineering and Materials Research Laboratory, University of Illinois Urbana-Champaign, Urbana, IL 61801, USA.

*Corresponding author. Email: chenli@ucr.edu (C.L.); zren@uh.edu (Z.R.); rwilson@ucr.edu (R.B.W.)

**Abstract:** Materials with high thermal conductivity (Λ) are needed to conduct heat away from hot spots in high power electronics and optoelectronic devices. Cubic boron arsenide (c-BAs) has a high thermal conductivity due to its special phonon dispersion relation. Previous experimental studies of c-BAs report a room-temperature thermal conductivity between 1000 and 1300 W m$^{-1}$ K$^{-1}$. We synthesized high purity isotopically enriched c-BAs single crystals with room-temperature thermal conductivity of ≈1500 W m$^{-1}$ K$^{-1}$. Using time-domain thermoreflectance (TDTR), we measured thermal conductivity and found a $1/T^2$ temperature dependence between 300 K and 600 K – slightly stronger than predictions from state-of-the-art theoretical models. Brillouin and Raman scattering revealed minimal changes in phonon frequencies over the same temperature range, suggesting that the observed $1/T^2$ dependence is not caused by temperature dependent changes in phonon dispersion. To probe defect densities in the BAs crystals we studied, we conducted transient reflectivity microscopy (TRM) measurements of absorption at sub-bandgap photon energies. We observe a correlation between TRM signal intensity and thermal conductivity. Notably, samples with thermal conductivity near 1500 W $m^{-1}$ $K^{-1}$ still exhibited nonzero TRM signals, suggesting the presence of defects despite the high thermal conductivity.

## Introduction

Cubic boron arsenide (c-BAs, for simplicity, BAs is used throughout the paper) is of interest for next generation electronics due to its combined high thermal conductivity (> 1000 W $m^{-1}$ $K^{-1}$) and high carrier mobility (> 1400 $cm^2$ $V^{-1}$ $s^{-1}$)[1–8]. BAs's exceptional transport properties were first predicted by first-principles calculations that considered the effect of its special phonon dispersion on phonon scattering[3,6,7,9]. The difference in atomic mass between B and As leads to a large acoustic-optical (*a-o*) phonon frequency gap. This large gap, together with bunching of acoustic phonon branches and unusual chemical bonding[9], leads to weak phonon-phonon scattering. Weak polarity and high frequency of optical phonons also suppress polar scattering of charge carriers[10]. As a result, both heat carriers (acoustic phonons) and charge carriers (electrons and holes) have long mean free paths.

First-principles calculations predict a room-temperature thermal conductivity of ≈ 1200 W $m^{-1}$ $K^{-1}$ for $^{nat}$BAs[3,11,12], and ≈ 1330 W $m^{-1}$ $K^{-1}$ for isotopically pure BAs[13]. These predictions are in reasonable agreement with prior experimental observations of Λ between 1000 and 1300 W $m^{-1}$ $K^{-1}$ for natural BAs[4,5], and between 1160 and 1260 W $m^{-1}$ $K^{-1}$ for isotopically enriched BAs[3,13,14]. (Early theoretical calculations predicted higher values of ≈ 1400 and 1700 W $m^{-1}$ $K^{-1}$ for natural and isotopically pure BAs, respectively[7,15]. Subsequent improvements in computational approach led to a ≈ 20% reduction in predictions for Λ[3,11–13].)

Theoretical predictions for the intrinsic Λ of BAs assume minimal defects. Recent studies suggest defect levels in BAs crystals studied to date are not negligible. Chen *et al.* used secondary ion mass spectroscopy and electron probe microanalysis to study impurities in BAs crystals[16]. The crystals were prepared in the same way as those studied in Ref.[3] and possessed room-temperature Λ ≈ 900 W $m^{-1}$ $K^{-1}$. They observed a Si impurity level of 0.047 at% ($3.4 \times 10^{19}$ $cm^{-3}$). BAs crystals are often reported to have peaks in their photoluminescence spectra near 1.5 eV[17,18]. Originally, these peaks were interpreted to imply a band gap of 1.5 eV in BAs[19,20].

However, recent studies have found the band gap of BAs to be between 1.8 and 2 eV[17,18,21–23], and credit photoluminescence peaks at 1.5 eV to group IV impurities[18].

Temperature-dependent thermal conductivity ($\Lambda$ *vs.* $T$) measurements can help identify the impact of defects on thermal transport. In insulating materials, $\Lambda$ *vs.* $T$ hinges on the temperature ($T$) dependence of phonon scattering rates. Phonon-defect scattering rates are independent of $T$[24,25], while phonon-phonon scattering rates are temperature sensitive[7]. So, $\Lambda$ is expected to vary less with temperature if defect concentrations are appreciable. Reports for $\Lambda$ *vs.* $T$ between 300 and 600 K in BAs vary between $\Lambda \propto 1/T^{1.3}$ and $1/T^{2}$[3–5]. First-principles calculations predict $\Lambda \propto 1 / T^{1.6}$ between 300 and 600 K for $^{nat}$BAs[3,7]. For isotopically pure BAs, its temperature dependence is predicted to fall between $\Lambda \propto 1/T^{1.7}$ and $1/T^{1.8}$[9,13,15]. Samples with higher defect concentrations are expected to have lower ambient thermal conductivity and weaker temperature dependence due to phonon-defect scattering. However, prior experimental results on BAs from different groups do not follow this trend[3–5]. For example, Kang *et al.* report a room-temperature thermal conductivity of 1300 W m$^{-1}$ K$^{-1}$ with a $\Lambda \propto 1/T^{1.3}$ dependence[4], while Tian *et al.* report a lower ambient thermal conductivity of 1160 W m$^{-1}$ K$^{-1}$ with $\Lambda \propto 1/T^{2}$[3]. This inconsistency in previous work — where expected trends between room-temperature thermal conductivity and temperature dependence are not observed — suggests potential systematic errors in measurement approaches that may complicate direct comparisons of thermal conductivity across studies. To address this issue, a single systematic study of $\Lambda$ versus $T$ in BAs crystals with varying impurity levels is necessary.

In this study, we synthesized high-quality, isotopically enriched cubic BAs single crystals exhibiting thermal conductivity as high as 1500 W m$^{-1}$ K$^{-1}$ – exceeding the theoretically predicted value of 1330 W m$^{-1}$ K$^{-1}$[13]. To investigate the temperature dependence of thermal conductivity between 300 and 600 K, we performed time-domain thermoreflectance (TDTR) measurements on a series of BAs samples with room-temperature thermal conductivities ranging from 700 to 1500 W m$^{-1}$ K$^{-1}$. As expected, samples with higher thermal conductivity at ambient conditions showed a stronger temperature dependence. In particular, the sample with 1500 W m$^{-1}$ K$^{-1}$ thermal conductivity followed a $\Lambda \propto 1/T^{2}$ trend over the measured temperature range – slightly stronger than the theoretically predicted dependence of $\Lambda \propto 1/T^{1.7}$ to $1/T^{1.8}$[9,13,15]. Furthermore, we introduced a qualitative method for distinguishing high-thermal-conductivity BAs using transient reflectivity microscopy with sub-bandgap photon excitation.

**Methods**

To study the temperature dependent thermal conductivity of isotopically enriched BAs, we synthesized a variety of BAs crystals by chemical vapor transport (CVT) as described previously[3,26], but with modifications (summarized below) that resulted in fewer defects and higher thermal conductivity.

The modifications in the CVT synthesis are 1) all source ingredients, isotopically enriched boron (B, 99.9999%), arsenic (As, 99.99999%), iodine ($I_2$, 99.9985%) (metal basis), from Alfa Aesar were further treated before sealing in the quartz tube to eliminate contaminants such as carbon (C), oxygen (O), sulfur (S), etc. that are all detrimental to the thermal conductivity. 2) The ratio between As and B was increased to a higher value than that in our prior work[3,26]. This was done to build a higher As partial pressure inside the quartz tube, and therefore decrease the As deficiency in BAs crystals. 3) The BAs crystals were removed from the tube in a purely mechanical way. Previously, our removal process involved soaking samples in aqua regia over 24 hours[3,26], and this etching process may have introduced point defects. Typical dimensions of our BAs samples are hundreds of microns in lateral dimensions and tens of microns in thickness.

Due to the small size of the synthesized BAs samples (hundreds of microns in lateral dimensions and tens of microns in thickness), traditional steady-state thermal conductivity measurements, which require larger sample sizes, are not feasible. Among various transient techniques, TDTR has proven to be one of the most reliable methods for measuring thermal conductivity in such small-scale samples[27].

Time-domain thermoreflectance is a well-established pump/probe technique for characterizing the thermal properties of materials[27]. In our TDTR measurements, we used a 783 nm laser wavelength, with pump beam modulated at 10.7 MHz. We used a laser spot size of ~15 μm in $1/e^2$ radius. In this configuration, TDTR measures the cross-plane thermal conductivity (normal to the surface).

In addition to standard TDTR measurements, we performed beam-offset TDTR measurements of the in-plane thermal conductivity (parallel to the surface). For beam-offset measurements, the sensitivity of $\Lambda_{\mathrm{BAs}}$ is maximized when the laser spot size closely matches the heat diffusion length[28]. The heat diffusion length is $d = \sqrt{\Lambda / \pi C f}$, where $f$ represents the pump modulation frequency. To enhance the sensitivity of $\Lambda_{\mathrm{BAs}}$, we deliberately chose laser spot sizes and modulation frequencies that align with the heat diffusion length in our beam offset measurements, as detailed in Table S1. Further elaboration on the principles underlying TDTR and beam-offset TDTR can be found in Refs.[27,29,30]. Additional specifics regarding our pump/probe system can be found in Ref.[31].

In a TDTR experiment, the signal of interest is the ratio of in-phase and out-of-phase thermoreflectance signals detected by an RF lock-in amplifier, $-V_{in} / V_{out}$. We collect and analyze this signal as a function of the delay time between pump and probe laser pulses. Data is analyzed with a multilayer thermal model of heat diffusion, yielding the thermal effusivity ($\sqrt{\Lambda C}$, with $\Lambda$ and $C$ being thermal conductivity and heat capacity, respectively) of BAs.

For beam-offset TDTR measurements[29,30], we measure the out-of-phase signal, $V_{out}$, as a function of offset distance between pump and probe beams. We fit the full width at half maximum (FWHM) of the $V_{out}$ signal with a heat diffusion model to determine the thermal diffusivity ($\Lambda/C$) of BAs. Utilizing known literature data on the heat capacity of BAs (Fig. S1), we derived thermal conductivity values via TDTR and beam-offset TDTR measurements. Details about the heat diffusion model can be found in Ref.[32]. Sensitivity analysis and thermal parameters used in our experiments can be found in Supplementary Materials[33].

In our experimental setup for TDTR and beam-offset TDTR measurements at temperatures exceeding 300 K, measurements were performed with the sample in a vacuum chamber at pressure ~ $10^{-3}$ Torr. We monitored the temperature of the vacuum chamber's heater stage using a thermocouple. The measured samples are attached to the stage (close to the thermocouple) using carbon paste.

We performed transient reflectivity microscopy (TRM) using the same pump-probe system as the TDTR experiments. The only difference is that for TRM measurements, the BAs was not coated with an aluminum transducer. For most TRM measurements, the laser wavelength is fixed at 783 nm. For the wavelength-dependent TRM, the laser wavelength was tuned from 690 to 980 nm. The pump modulation frequency was 10.7 MHz, the incident pump power was 20 mW, and the $1/e^2$ laser radius was set to either 7.5 μm or 15 μm. We observed the $V_{in}$ signal was linearly proportional to laser power density. Therefore, for measurements with 7.5 μm laser radius, we normalized the $V_{in}$ signal by dividing it by 4. The observed $V_{in}$ signal is normalized by the power of pump beam and the voltage on the photodiode detector.

For wavelength-dependent TRM measurements, we use a Pt thin film as the control sample to calibrate the wavelength-dependent factors originating from optics. The measured thermoreflectance signal is

$$\Delta R = \frac{dR}{dT}\Delta T = C(\lambda) \cdot \frac{dR}{dT}(\lambda) \cdot [1 - R(\lambda)] \cdot P_{pump}(\lambda) \cdot V_{DC}(\lambda). \quad (1)$$

Here, $dR/dT$ is the thermoreflectance, $\Delta T$ is the temperature rise due to the heating of pump beam, $R$ is the reflectance, $P_{pump}$ is the pump power, and $V_{DC}$ is the voltage on the photodiode detector, which depends on the probe power and detector responsivity. $C(\lambda)$ is a wavelength-dependent factor related to optics in our pump/probe system. For example, the transmission of objective lens and the modulation envelope of electro-optic modulator are wavelength dependent. To calibrate $C(\lambda)$, we performed TRM on a Pt thin film as a function of wavelength. We measured $P_{pump}$ and $V_{DC}$ at every wavelength, and obtained wavelength dependent $(1 - R)$ and $dR/dT$ of Pt from Ref.[34]. We used this calibrated $C(\lambda)$ for wavelength-dependent TRM on BAs samples.

In addition to TDTR and TRM measurements, we also performed Raman scattering, Brillouin-Mandelstam spectroscopy (BMS)[35], forced Brillouin scattering, and photoluminescence (PL) on our BAs samples. Details about these measurements can be found in Supplementary Materials[33].

## Results and Discussion

We characterized our BAs samples using X-ray diffraction (XRD), Raman scattering, photoluminescence (PL), and Brillouin-Mandelstam spectroscopy[35] see Fig. 1. The XRD results were consistent with a (111) growth facet. In most samples, we observed PL spectra centered near 1.75 eV. But some samples exhibited additional PL peaks centered at lower energy. More PL results can be found in Fig. S2. Overall, our PL results were consistent with the spectra observed in prior studies[1,2,17,18]. We did not observe a clear correlation between thermal conductivity and PL spectra. Some BAs crystals with high thermal conductivity, e.g. above 1400 W $m^{-1}$ $K^{-1}$, had peaks near 1.5 eV, which indicated the presence of color centers. Alternatively, some BAs crystals with lower thermal conductivity, e.g. < 1000 W $m^{-1}$ $K^{-1}$, showed no PL peaks in the near-infrared, like we would expect for defect free crystals.

We observed Raman spectra that are consistent with prior observations for BAs crystals with similar isotope concentrations[13,36]. We observe a peak near 700 $cm^{-1}$, like has been reported for isotopically pure $^{11}$BAs. For BAs with a natural isotope abundance, we would expect peaks at both 700 and 720 $cm^{-1}$. Li *et al.*[5] observed a relationship between thermal conductivity and the integration of the Raman intensity from 1050 to 1150 $cm^{-1}$ in BAs. We analyzed our Raman spectra following their procedure and observed a similar but less sharp trend, see Fig. S3b. To analyze the background of the Raman spectra, we subtract the spectra of BAs by the dark background of the monochromator, *i.e.*, the signal collected at the same collection time without any incident light.

Our BMS measurements revealed TA and LA sound velocities of 4800, 5080, and 8320 m/s along the direction of the refracted laser beam, which is ~ 8° from the [111] direction. In the [111] direction, we expect the transverse branches to degenerate. The splitting of the transverse peaks indicates a small misalignment of the measurement axis relative to the [111] direction. The angle between the incident laser beam and surface normal ([111]) is 30°. The calculated refraction angle for the 30-degree incidence is 8°. Therefore, we are measuring the sound velocity along a direction which is 8° away from [111]. Our BMS results are consistent with prior experimental results[37] of 4978 and 8513 m/s for the transverse and longitudinal acoustic phonons along [111] direction, respectively, obtained using picosecond interferometry. BMS results on BAs samples with various thermal conductivity can be found in Table. S2.

Surface roughness can negatively affect the reliability of TDTR and beam-offset TDTR measurements by introducing unwanted modulation of diffusely scattered laser light by thermoelastic effects[38,39]. To confirm our high-thermal-conductivity BAs samples have smooth surfaces, we performed atomic force microscopy (AFM) on an area of $10 \times 10\ \mu m^2$ of a BAs sample with $\Lambda_{300K}$ of 1500 W $m^{-1}$ $K^{-1}$, see Fig. 2a. The measured root-mean-square (RMS) roughness is 1.8 nm, which is more than smooth enough for TDTR measurements to be reliable. Picosecond acoustic echoes in TDTR measurements can deliver some indirect information of interface roughness between Al and substrates[40]. Substrates with smooth surfaces yield sharp and narrow acoustic echoes in $V_{in}$ signals. Sometimes, acoustics from rough surfaces yield wider acoustic echoes, which is an indicator that the surface roughness is too high. We compared the acoustic echo from the Al/BAs sample with echoes from commercially available Si and diamond wafers coated with Al (Fig. 2b). We observed that the acoustic echoes from the BAs sample have similar widths as echoes from Si and diamond wafers. For most BAs samples, we did not perform AFM measurements. Instead, we checked all picosecond acoustic peaks and confirmed that the specular reflectance of all samples was within 5% of the expected value for an Al-coated surface.

Time-domain thermoreflectance (TDTR) measurements reveal the ambient thermal conductivity ($\Lambda_{300K}$) of the $^{11}$B isotopically enriched BAs crystals are as high as 1500 W m$^{-1}$ K$^{-1}$, higher than previous experimental observations[3–5]. As part of this study, we measured the thermal conductivity of more than fifty BAs crystals. Among these, four crystals had thermal conductivity above 1400 W m$^{-1}$ K$^{-1}$. TDTR data for one of the high-thermal-conductivity crystals (sample BAs-1500) is shown in Fig. 3(a). TDTR data for the other three are shown in Fig. S4. Sensitivity analysis of TDTR measurements can be found in Fig. S5. Thermal conductivity of all measured samples is shown in Table S3.

We observe that BAs crystals with $\Lambda_{300\mathrm{K}} = 1500$ W m$^{-1}$ K$^{-1}$ exhibit $\Lambda \propto 1 / T^2$ between 300 and 600 K (Fig. 3(b)). This is a slightly stronger temperature dependence than being predicted by first-principles calculations[3,7,9,13,15], and stronger than prior experimental results of BAs[4,5]. We corroborated the observed $1/T^2$ dependence by performing separate beam-offset TDTR measurements[30] of the in-plane thermal conductivity versus temperature on BAs-1500, see Figs. 5 and 6(a). Additionally, we measured another BAs sample with $\Lambda_{300\mathrm{K}} = 1500$ W m$^{-1}$ K$^{-1}$ (sample B7N2) and observed the same temperature dependence, see Fig. 6(a). Further sanity checks of our TDTR measurements and analysis can be found in Figs. S6 – S8.

To investigate the mean-free-path distribution of heat-carrying phonons in BAs, we conducted TDTR measurements as a function of laser spot size[41–43]. When the laser spot size is small, the in-plane heat current carried by phonons with long mean free path is less than what Fourier's law predicts[41]. This results in spot-size dependent apparent thermal conductivity ($\Lambda_A$), where we define $\Lambda_A$ as the $\Lambda$ that yields the best fit of the heat diffusion model to the experimental data. We measured $\Lambda_A$ of the BAs-1500 sample at temperatures of 300, 450, and 600 K (Fig. 3(c)). At 300 K, a spot-size reduction from 15 to 1.7 µm led to a ~ 20% reduction in $\Lambda_A$. A similar suppression of $\Lambda_A$ was observed in silicon and diamond[41–43]. The 20% drop we observed is much smaller than prior reports for $\Lambda_A$ versus spot size in BAs[4,5]. A weak dependence of $\Lambda_A$ on spot size for BAs is in qualitative agreement with first-principles calculations[3], and consistent with Peierls-Boltzmann transport equation simulations of non-diffusive heat transfer in TDTR measurements of BAs[44]. First-principles calculations predict 80% of the heat is carried by phonons with mean free paths between 0.3 and 1.5 µm. At 450 and 600 K, we observe that laser spot size does not affect $\Lambda_A$ (Fig. 3(c)). This suggests that for $T \geq 450$ K, the mean free paths of all heat-carrying phonons are less than 1.7 µm. To minimize mean-free-path effects in the rest of our experiments, we used a $1/e^2$ laser radius greater than 10 µm. Additionally, all samples in this study have a thickness greater than 20 µm, and we employed a modulation frequency of 10.7 MHz, corresponding to a thermal penetration depth of 4.6 µm. As a result, mean free path related effects should be minimal in our measurements.

To further explore the temperature-dependent thermal conductivity of BAs, we selected four other BAs samples and measured their $\Lambda$ *vs.* $T$, see Fig. 4. These samples had $\Lambda_{300K}$ of 1350, 1200, 1000, and 700 W m$^{-1}$ K$^{-1}$. We also measured a type IIa diamond crystal from Element Six as a control sample. We fit the measured data of $\Lambda$ *vs.* $T$ by $\Lambda \propto 1/T^{\alpha}$ to obtain the temperature exponent α. Details about the power law fitting can be found in Supplementary Materials[33]. We plot α *vs.* $\Lambda_{300K}$ for the selected BAs samples (Fig. 4(b)). As $\Lambda_{300K}$ decreases, the temperature dependence weakens. As described above, this is expected since phonon-defect scattering lowers α. The temperature exponent of diamond is 1.2, in good agreement with prior reports[4,7]. Our BAs sample with $\Lambda_{300K}$ = 1000 W m$^{-1}$ K$^{-1}$ has similar temperature dependence as the results reported in Ref.[5]. A plausible explanation for the inconsistency in trends observed in prior studies between $\Lambda_{300K}$ and α is that systematic errors differ across studies, complicating direct comparisons.

The temperature dependence of $\Lambda$ we observe for BAs is unusually large compared to other semiconductors. At high temperatures, the temperature dependence of $\Lambda$ is primarily determined by the type of scattering processes that limit mean free paths of heat carrying phonons. In most high purity single crystals, three-phonon scattering is the dominant process, resulting in $\Lambda \propto 1/T$ ($\alpha = 1$)[7,9]. Three-phonon-scattering rates increase linearly with temperature because the scattering rate is approximately proportional to the number of phonons per unit volume, a value that scales linearly with temperature at high $T$. A higher temperature exponent can indicate higher-order phonon scattering processes[9]. Four-phonon-scattering rates are expected to increase quadratically with temperature because, such scattering processes are approximately proportional to the square of the phonon population at high $T$[7,45]. Note that the Debye temperature of BAs is ~ 670 K[46], comparable to other III-V semiconductors[47,48]. Similar Debye temperature indicates the strong temperature dependence is not a result of heat capacity change. Electron-phonon scattering is another possible scattering channel. However, the predicted thermal conductivity of BAs displays a weaker ($1/T$) temperature dependence when adding electron-phonon interactions to three- and four-phonon interactions[49].

In the limit where four-phonon scattering processes dominate, α is expected to be close to but smaller than 2 in defect-free samples[7]. Some materials exhibit $\alpha > 1$, but it is rare for α being close to 2. For instance, BP and InP have temperature exponents of 1.4[50] and 1.5[51], respectively. BP and InP are III-V semiconductors with an *a-o* gap due to the large mass ratio between constituent atoms[6,9]. A large *a-o* gap limits the phase space of three-phonon scattering, which can increase the importance of four-phonon scattering processes[9]. BP, InP, and BAs are examples of this trend. The relationship between α and mass ratio for III-V compounds with zincblende structure is summarized in Fig. S9.

In BAs, four-phonon-scattering processes are expected to be important, but not dominant[7,9,15]. For isotopically pure BAs, theoretical predictions for α lie between 1.7 and 1.8[9,13,15]. Overall, the theoretical predictions agree well with our measurements in Fig. 3(b), and their difference is within 15% at all temperatures. However, we note that the value for α we experimentally observe is 20% higher than theoretical prediction, and this minor disagreement merits some discussion.

The difference in experimental vs. theoretical α could originate from several factors. The ratio of four-phonon to three-phonon scattering rates in BAs could be higher than predictions[9,13,15]. Although first-principles calculations do not use fitting parameters, their accuracy is affected by a variety of factors[11,52]. Jain and McGaughey found that the choice of exchange-correlation function can change the predicted thermal conductivity of Si by ~ 30%[52]. Zhou *et al*. report that fourth-order force constants are sensitive to the energy surface roughness of the exchange correlation functionals[11].

First principles calculations predict $\Lambda_{300K}$ of 3100 W m$^{-1}$ K$^{-1}$ and $\alpha = 0.9$ if only considering three-phonon scattering[13]. In the limit that four-phonon scattering is the only dominant scattering process, $\Lambda_{300K}$ is expected to be an order of magnitude higher than what we observe, i.e., $1.5\times10^4$ W m$^{-1}$ K$^{-1}$ with $\alpha = 2$ [7,53]. Therefore, it is not plausible that only four phonon processes matter. Another possible explanation for $\alpha \geq 2$ is that higher-order processes than four-phonon scattering affect $\Lambda$ in BAs. From the perspective of relaxation time approximation, adding higher-order processes to existing first-principles calculations would increase $\alpha$, but it would also lower predictions for $\Lambda_{300K}$. Ravichandran and Broido[9] calculated the four-phonon scattering rates of various III-V semiconductors including BAs and observed similar four-phonon scattering rates and lack of selection rules affecting them. Therefore, Ravichandran and Broido expect negligible effect of higher-order processes on the thermal conductivity of III-V semiconductors. However, detailed theoretical calculations including higher-order than four-phonon processes have not yet been carefully explored. Higher order processes being important could explain our results if three- and four-phonon scattering processes were a bit weaker than currently predicted by first-principles theory.

Another possible explanation for why $\alpha$ in BAs is higher than theoretical predictions is temperature-induced changes of the phonon dispersion. Three- and four-phonon scattering rates in BAs are sensitive to the *a-o* gap and the bunching of acoustic modes[6,53–55], Therefore, temperature-induced changes in phonon dispersion would strongly impact the $\Lambda$. We do not find this hypothesis likely for several reasons. Although there are no experimental measurements of phonon dispersion versus temperature, DFT calculations suggest that phonon dispersion is only weakly temperature-dependent[56]. This is supported by our temperature-dependent Brillouin and Raman scattering experiments, which show minimal changes in phonon frequencies with increasing temperature (< 1% from 300 to 600 K, see Fig. 7 and 8). Therefore, it is unlikely that temperature-induced changes to the phonon dispersion can explain why $\Lambda$ is proportional to $1/T^2$. (Raman and Brillouin scattering do not measure the high-frequency acoustic phonons that are most responsible for heat transfer in BAs. A measurement of the full phonon dispersion spectrum at high temperature is necessary to completely rule out this hypothesis.)

To characterize defects in our samples, we performed pump/probe transient reflectivity microscopy on bare BAs samples with incident photon energy of 1.58 eV (Fig. 9). In these experiments, we irradiated the BAs surface with a pump beam. We measured the intensity of a reflected probe beam as a function of pump/probe delay time. The laser energy is ~ 0.26 eV less than the 1.84 eV band gap of BAs[17,23]. Therefore, in the absence of impurities, we expect negligible absorption and no pump-induced change in the reflectance of BAs. In the presence of impurities that form defect states in the band gap, we expect measurable TRM signal. The effect of impurity on absorption will depend on the energy level of the defect state. If the defect state's energy is near the conduction or valence band edges (ionization energy $\leq 3k_BT \approx$ 75 meV), the impurity affects absorption by introducing free carriers, see Fig. 9(a). Alternatively, defects that form states far from the conduction or valence band edges will allow optical transitions to (or from) the defect state, see Fig. 9(b). For impurities with moderate ionization energies, *e.g.*, 100 - 300 meV, we expect both these absorption mechanisms will matter.

We observed non-zero TRM signals in most BAs samples, including the ones with $\Lambda_{300\mathrm{K}} =$ 1500 W m$^{-1}$ K$^{-1}$ (Fig. 9(c)). We performed TRM measurements at $\geq 10$ spots on each sample. Most measurements were performed with a pump fluence of 0.7 J/m$^2$ and probe power of 2 mW. A few measurements were conducted with different pump fluences and probe powers. For these measurements, to facilitate comparisons in Fig. 9, we scaled the signals by a factor to account for the difference in pump fluence. Each marker in Fig. 9(c) and 9(f) represents the average value of all measured spots, while error bars represent the standard deviation. The data in Fig. 9(d) and 9(e) are for individual spots, not averages. We observed a rough correlation between $\Lambda_{300\mathrm{K}}$ and TRM signals (Fig. 9(c)). The reason why the correlation is not very clear may originate from the different penetration depth between TRM and thermal conductivity measurements. The optical penetration depth of TRM measurement is on the order of tens of μm, while the thermal penetration depth of TDTR measurements is about 4 μm. We observed that the as-grown BAs samples showed different thermal conductivity at top and bottom surfaces (see Table S3). This indicates the two different sides of one single BAs sample have different defect concentrations. We performed wavelength-dependent pump/probe TRM measurements (Fig. 9(e)) on BAs-1500 as well as two BAs samples whose $\Lambda_{300K}$ are 1000 (BAs-1000) and 400 W m$^{-1}$ K$^{-1}$ (BAs-400), respectively. For these measurements, we fixed our laser beam on a region that had a small TRM signal, i.e., a region with a low density of defects. We observed an increasing TRM signal as photon energy approaches the band gap of BAs. As a set of control experiments, we also performed TRM measurements with a laser energy of 1.58 eV on crystals with different band gaps: Si (1.12 eV)[57], GaAs (1.42 eV)[58], GaP (2.26 eV)[59], and GaN (3.39 eV)[60]. The carrier densities of the GaP and GaN samples are $(4{\sim}6) \times 10^{16}$ cm$^{-3}$ and $\sim 1 \times 10^{18}$ cm$^{-3}$, respectively. As expected, TRM signals are large for crystals with band gaps less than 1.58 eV, and negligibly small for crystals with band gaps greater than 1.58 eV, see Fig. 9(f). We also measured intentionally doped n-type and p-type GaP crystals with specified carrier concentrations on the order of $10^{18}$ cm$^{-3}$. The TRM signal from BAs-1500 is the same order of magnitude as the TRM signal from GaP with $\approx 10^{18}$ cm$^{-3}$ p-type Zn impurities, see Fig. S10. We note that the TRM measurements are only qualitative, and do not provide a quantitative measure of defect concentration.

To explore sample homogeneity, we mapped how TDTR and TRM signals change with position for two of the crystals: BAs-1500 and BAs-1000, see Fig. 10. The TRM and TDTR maps are not particularly well correlated: the TRM maps display bright spots that are not visible on the TDTR maps. Overall, the TDTR measurements indicate fairly uniform thermal conductivity, with variations within $\pm$ 20% across most areas. In contrast, the TRM maps show variations spanning orders of magnitude. Moreover, we observed nearly zero TRM signals on a BAs sample (Fig. S10 and S11) with $\Lambda_{300K} \approx 800$ W m$^{-1}$ K$^{-1}$. We interpret these observations to mean that point defects within the sample can affect the absorption/emission spectrum while having little impact on thermal conductivity, and vice-versa. As we explain below, this explanation aligns with prior theoretical modeling of how point defects affect optical and thermal properties[18,61]. We also note TDTR is local to within 5 to 10 microns of the surface, and TRM is not. Finally, it is interesting that despite the high purity of these crystals, there is still significant spatial inhomogeneity in TRM maps, and therefore inhomogeneity in defect concentration.

Consideration of how defects affect thermal conductivity vs. optical properties is useful when comparing TRM and thermal conductivity results. Prior studies show the presence of impurities such as C, Si, O, and I in BAs crystals[3,16,18,62]. Theoretical calculations[63] predict that the substitutional defects formed by O on As sites ($O_{As}$) lead to defect states in the middle of the band gap. Alternatively, C and Si impurities on As sites ($C_{As}$ and $Si_{As}$) are predicted to be shallow acceptors with ionization energies of 0.08 and 0.07 eV, respectively[63]. So, $C_{As}$ and $Si_{As}$ are expected to affect absorption by introducing free holes. C and Si impurities on B sites ($C_B$ and $C_{Si}$) are predicted to be donors with ionization energies of 0.28 and 0.14 eV, respectively[63]. Therefore, $C_B$ and $Si_B$ are expected to affect absorption in one of the two ways mentioned above. Ionized $C_B$ and $Si_B$ impurities introduce free electrons, while $C_B$ and $Si_B$ that are not ionized will allow optical transitions from defect states into the conduction band.

All these types of impurities are also expected to have different effects on the thermal conductivity of BAs. Chen *et al.*[16] predict that reducing BAs's thermal conductivity by 10% requires a concentration of $\approx 3\times10^{18}$ cm$^{-3}$ neutral $C_B$ defects, $\approx 10^{18}$ cm$^{-3}$ $Si_{As}$ or $C_{As}$ defects, and $3\times10^{17}$ cm$^{-3}$ $O_{As}$ defects. So, in short, while impurities affect both thermal transport and below-band-gap optical absorption, we do not expect a perfect correlation between these physical properties.

Based on the TRM results, BAs samples with $\Lambda_{300K}$ of 1500 W m$^{-1}$ K$^{-1}$ are not entirely defect-free, although we cannot quantify the exact value. If the defect concentration in our BAs samples is high enough, e.g. $> 10^{18}$ cm$^{-3}$, the intrinsic thermal conductivity of BAs may be larger than the 1500 W m$^{-1}$ K$^{-1}$ value we report here, and the intrinsic temperature dependence between 300 and 600 K of BAs may be stronger than $1/T^2$.

In summary, we have observed thermal conductivity in high purity isotopically enriched BAs as high as 1500 W m$^{-1}$ K$^{-1}$. Crystals with high thermal conductivity have a strong temperature dependence ($\Lambda \propto 1 / T^2$). Measurements indicate that even the highest purity BAs crystals we studied have a high enough concentration of defects to affect optical properties in the near-infrared (photoluminescence, TRM signals). Whether the defect concentrations are high enough to impede thermal transport is not known. Our experimental observations for the thermal conductivity of BAs are in good agreement with state-of-the-art theory, but there is minor disagreement regarding the dependence of $\Lambda$ vs. $T$ that merits some additional consideration.

**Acknowledgments:**

R. B. W. acknowledges ULTRA, an Energy Frontier Research Center funded by the U.S. Department of Energy (DOE), Office of Science, Basic Energy Sciences (BES), Award #DE-SC0021230. C. L. acknowledges National Science Foundation (NSF) Award #1750786. A. A. B and F. K. acknowledge National Science Foundation (NSF) *via* a Major Research Instrumentation (MRI) project DMR 2019056

**Author contributions:**

S.H. and F.P. contributed equally to this work.

R.B.W. and Z.R. conceptualized this work.

F.P. and Z.R. performed crystal growth. F.P., G.A.G, H.S., B.A.N., and Z.R. developed the crystal growth and harvesting method.

S.H., R.B.W., F.A., and D.G.C. conducted time-domain thermoreflectance measurements. S.H. and R.B.W. conducted beam offset measurements, transient reflectivity microscopy. S. H. performed Raman scattering, photoluminescence, and forced Brillouin scattering measurements. F.P. performed X-ray diffraction, X.S. performed atomic force microscopy. Z.E., F.K., and A.B. performed Brillouin-Mandelstam spectroscopy scattering.

S.H. and R.B.W. performed thermal modelling.

S.H. performed visualization

R.B.W., C.L., and Z.R. acquired funding and supervised this work.

S.H. and R.B.W. wrote the original draft of the manuscript. All authors edited the manuscript.

**Competing interests:** Authors declare that they have no competing interests.

**Data availability:** The data that support the findings of this article are not publicly available upon publication because it is not technically feasible and/or the cost of preparing, depositing, and hosting the data would be prohibitive within the terms of this research project. The data are available from the authors upon reasonable request.

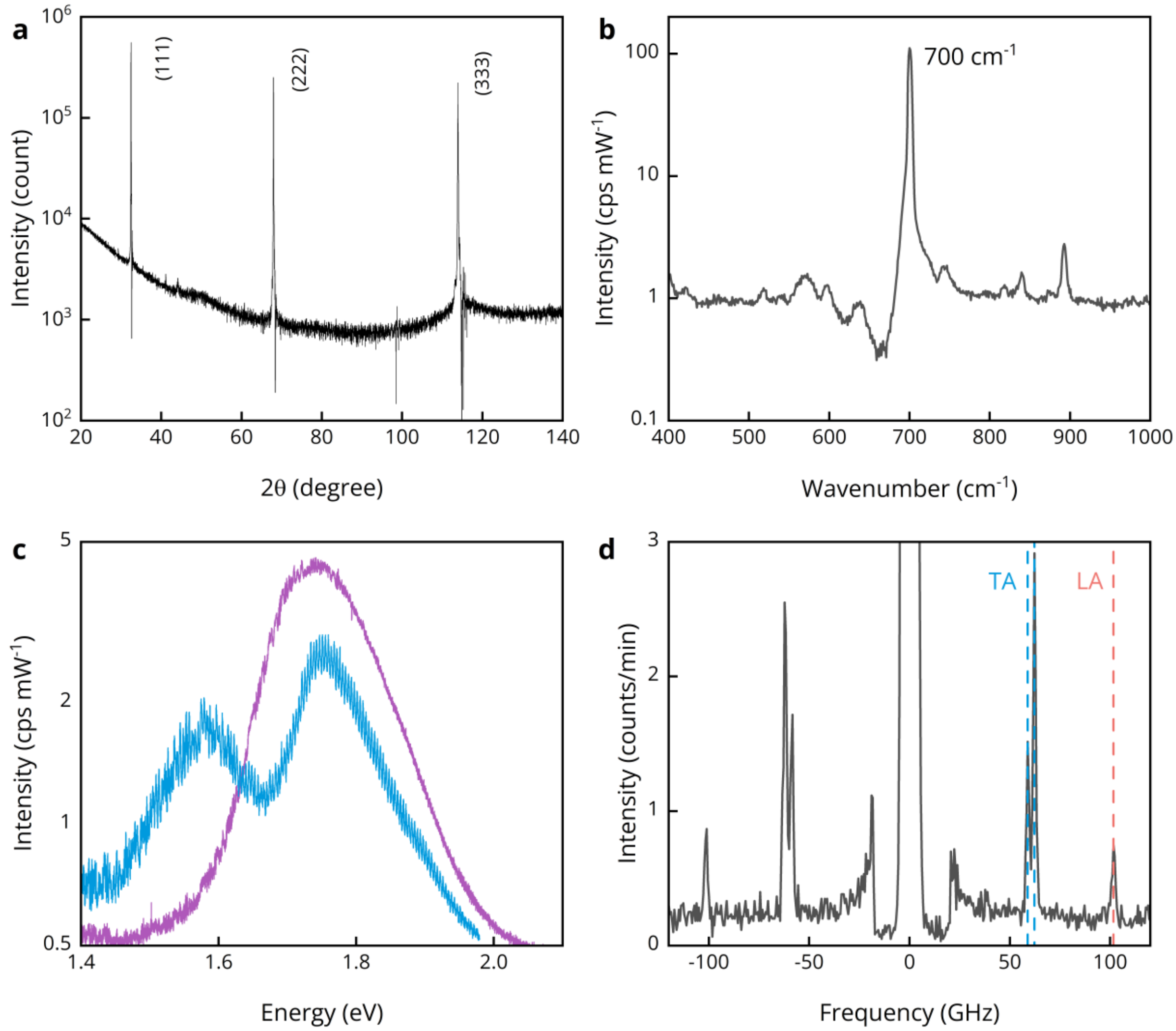

**Fig. 1.** (a) X-ray diffraction on a BAs sample with $\Lambda_{300K} = 1500$ W m$^{-1}$ K$^{-1}$. (b) Raman scattering on a $^{11}$B enriched BAs sample. (c) Photoluminescence spectra of BAs. The blue and purple curves are obtained from two different BAs samples. Both samples exhibit $\Lambda_{300K} = 1500$ W m$^{-1}$ K$^{-1}$. (d) Brillouin-Mandelstam spectrum of BAs. The two blue dashed lines mark the frequencies of the transverse acoustic phonon modes, while the pink dashed line indicates the frequency of the longitudinal acoustic phonon mode. All the measurements were conducted at room temperature. The unit 'cps' stands for counts per second.

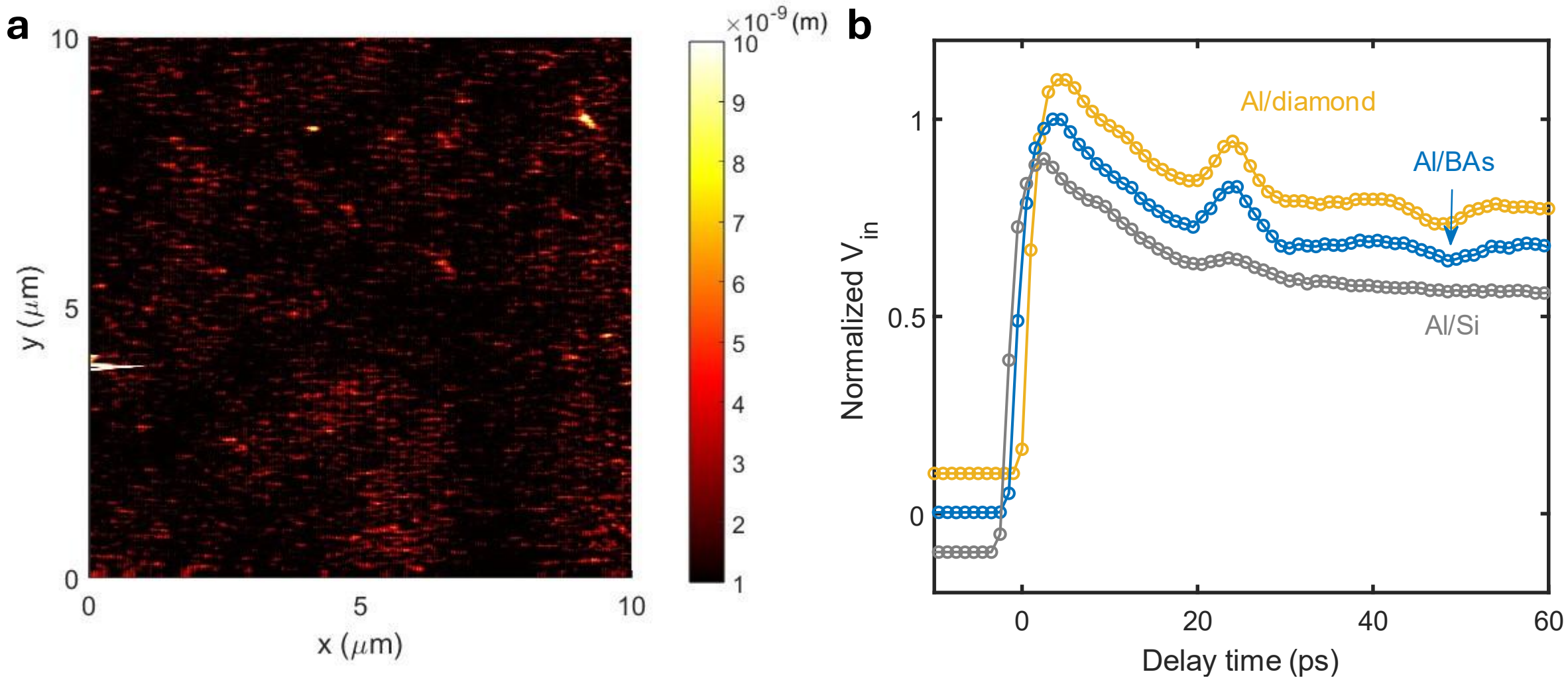

**Fig. 2.** Roughness characterization of a BAs sample with thermal conductivity of 1500 W $m^{-1}$ $K^{-1}$. (a) Atomic force microscopy on the BAs sample. The root-mean-square roughness of the measured area is 1.8 nm. (b) Comparison of the acoustic echoes between Al/BAs, Al/Si, and Al/diamond samples. The $V_{in}$ signal is normalized by its maximum value and shifted for clarity.

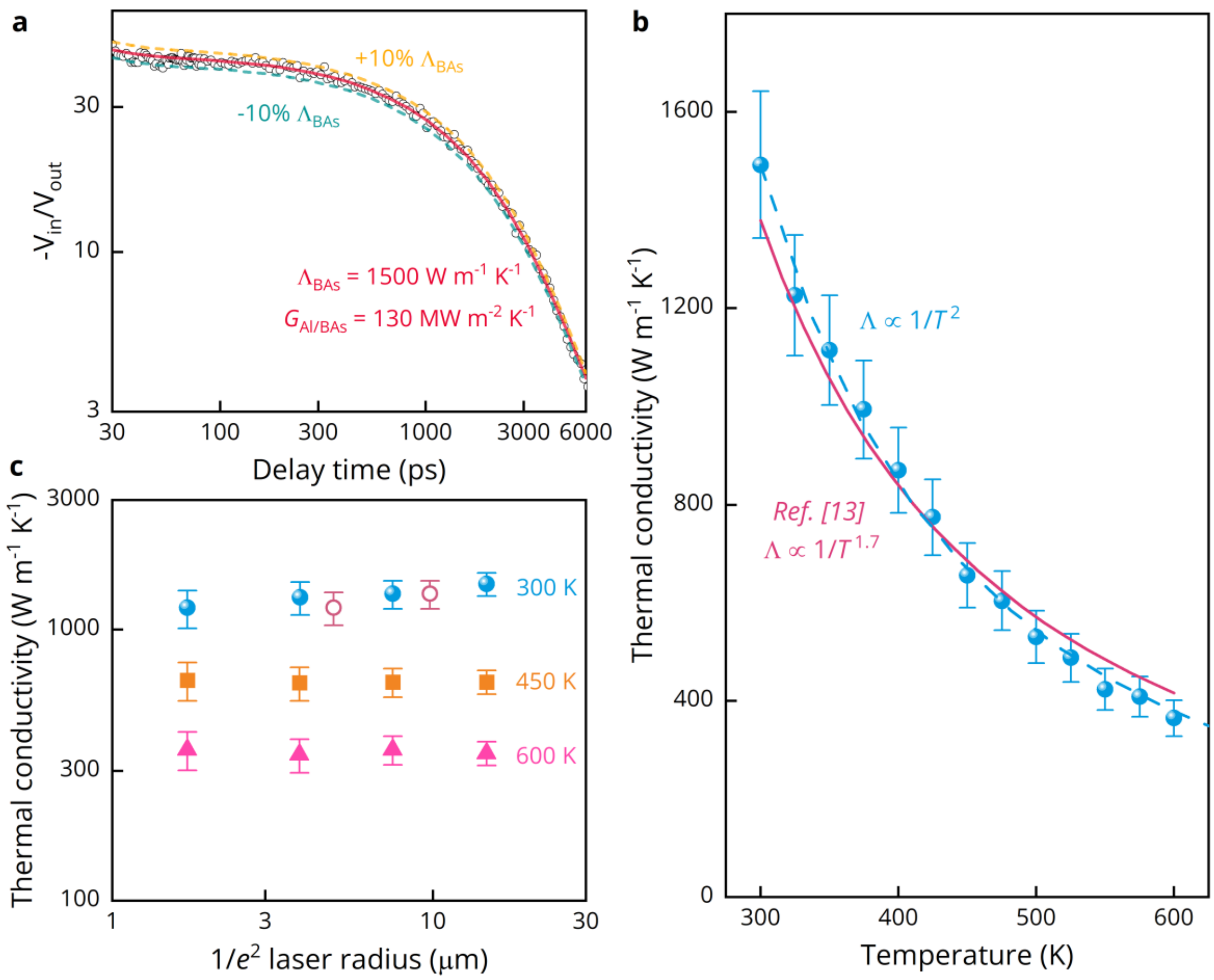

**Fig. 3.** (a) Time domain thermoreflectance data with the best fit of thermal model (red line) for a high purity BAs crystal with $\Lambda \approx 1500$ W m$^{-1}$ K$^{-1}$. The interface conductance between Al thin-film transducer and BAs crystal, which is also a fit parameter in the thermal model, is 130 MW m$^{-2}$ K$^{-1}$. The temperature-dependent interface conductance is shown in Fig. S1b. (b) BAs thermal conductivity versus temperature (markers), along with theoretical predictions for isotopically pure BAs[13]. The blue dashed line is a $\Lambda \propto 1/T^2$ fit to our data. (c) Apparent thermal conductivity of BAs versus laser spot size ($1/e^2$ radius) at 300, 450, and 600 K. All the measurements are conducted on one single sample. (Filled and open markers denote measurements performed at UCR and UIUC, respectively.)

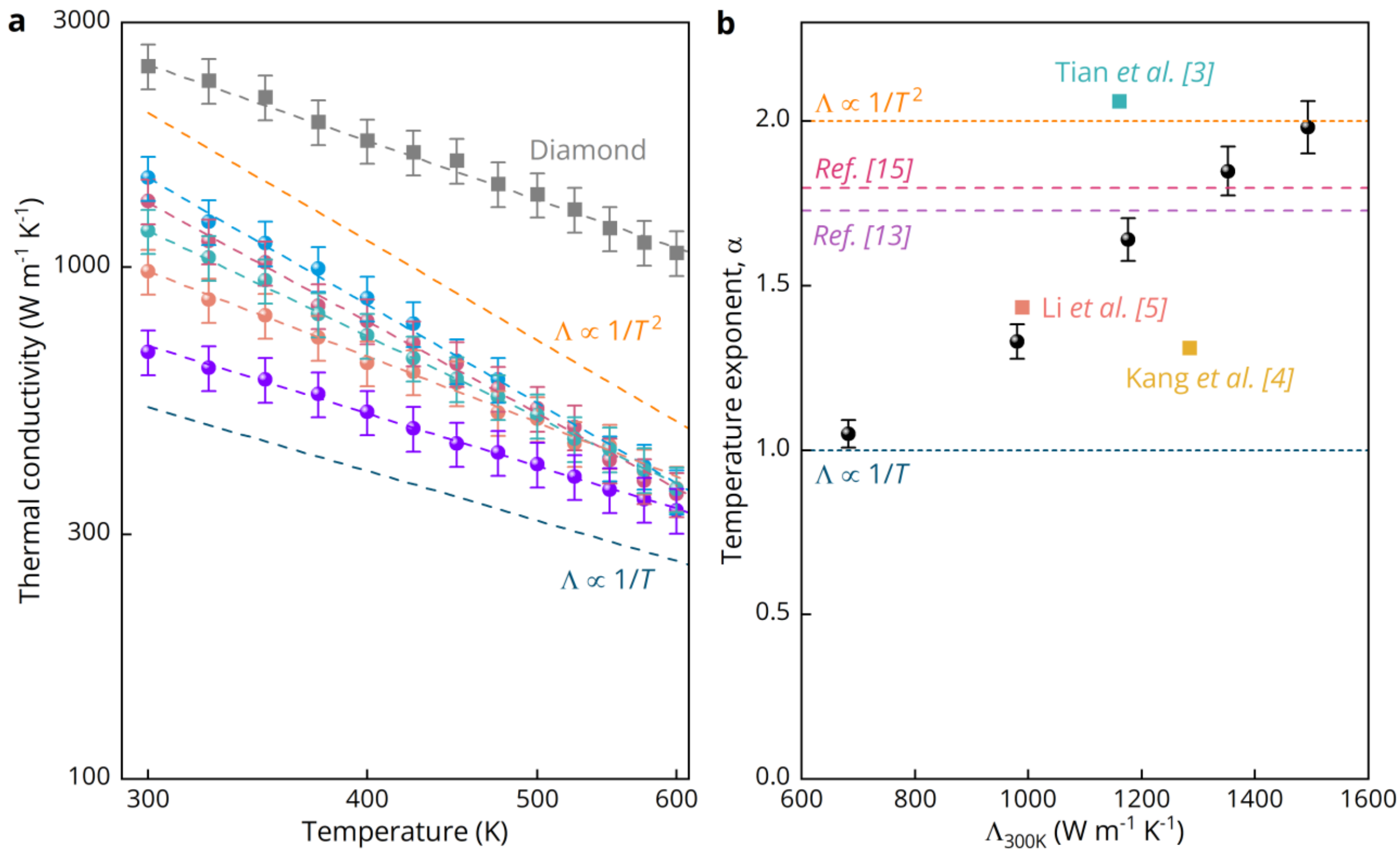

**Fig. 4.** (a) Temperature dependent thermal conductivity of five selected BAs crystals and a type IIA diamond measured using TDTR. (b) The relationship between the temperature exponent and the room-temperature thermal conductivity of the five BAs crystals shown in (a). Square markers are prior experimental results[3–5]. Pink[15] and purple[13] dashed lines are theoretical predictions of α for isotopically pure BAs.

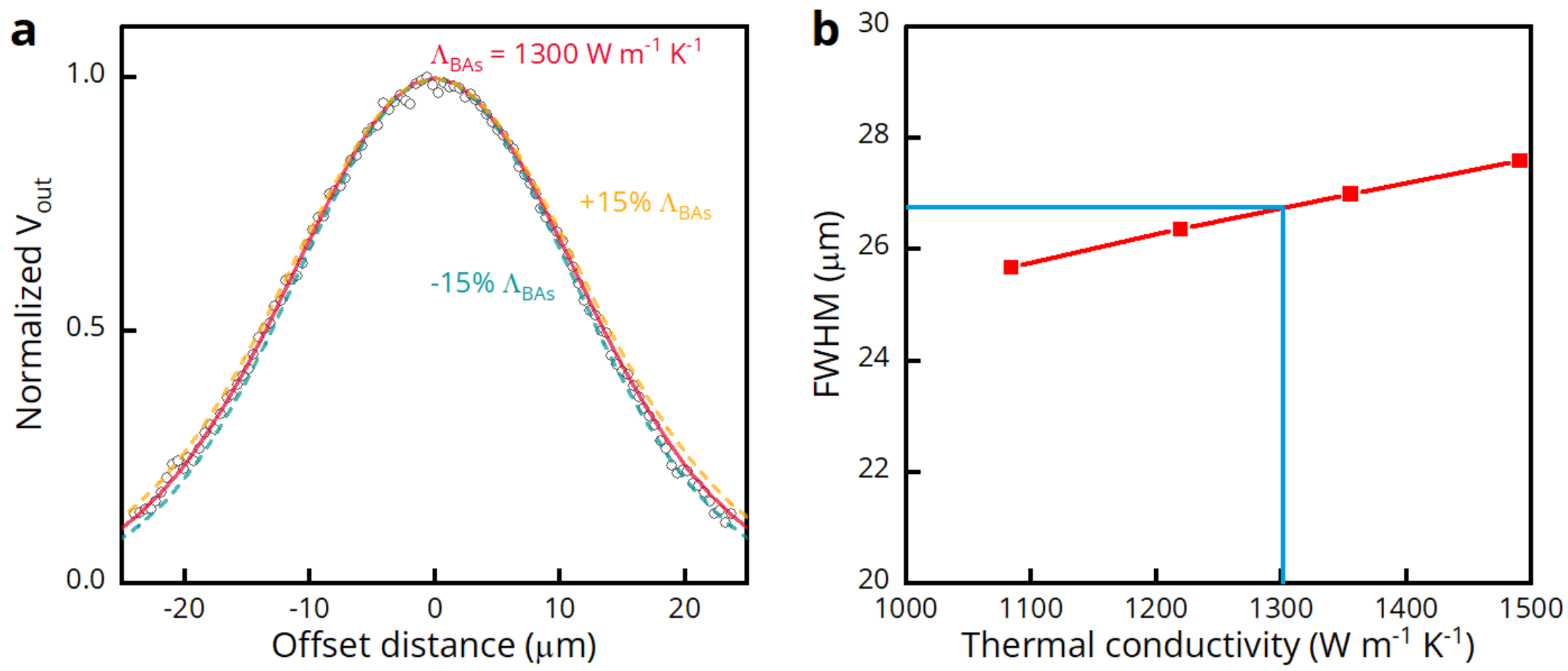

**Fig. 5**. Beam-offset TDTR measurements conducted on the BAs-1500 sample. (a) A beam offset scan (circles) along with the best fit (red line). Predictions for 15% larger and smaller thermal conductivity of BAs are shown as yellow and green dashed lines, respectively. The vertical axis is the normalized out-of-phase signal, $V_{out} \;/\; V_{out}(max)$. The horizontal axis represents the offset distance between pump and probe beams. (b) The full width at half maximum versus thermal conductivity predicted from the heat diffusion model. The horizontal blue line indicates the FWHM of the experimental data in panel (A), and the vertical blue line represents the fitted thermal conductivity.

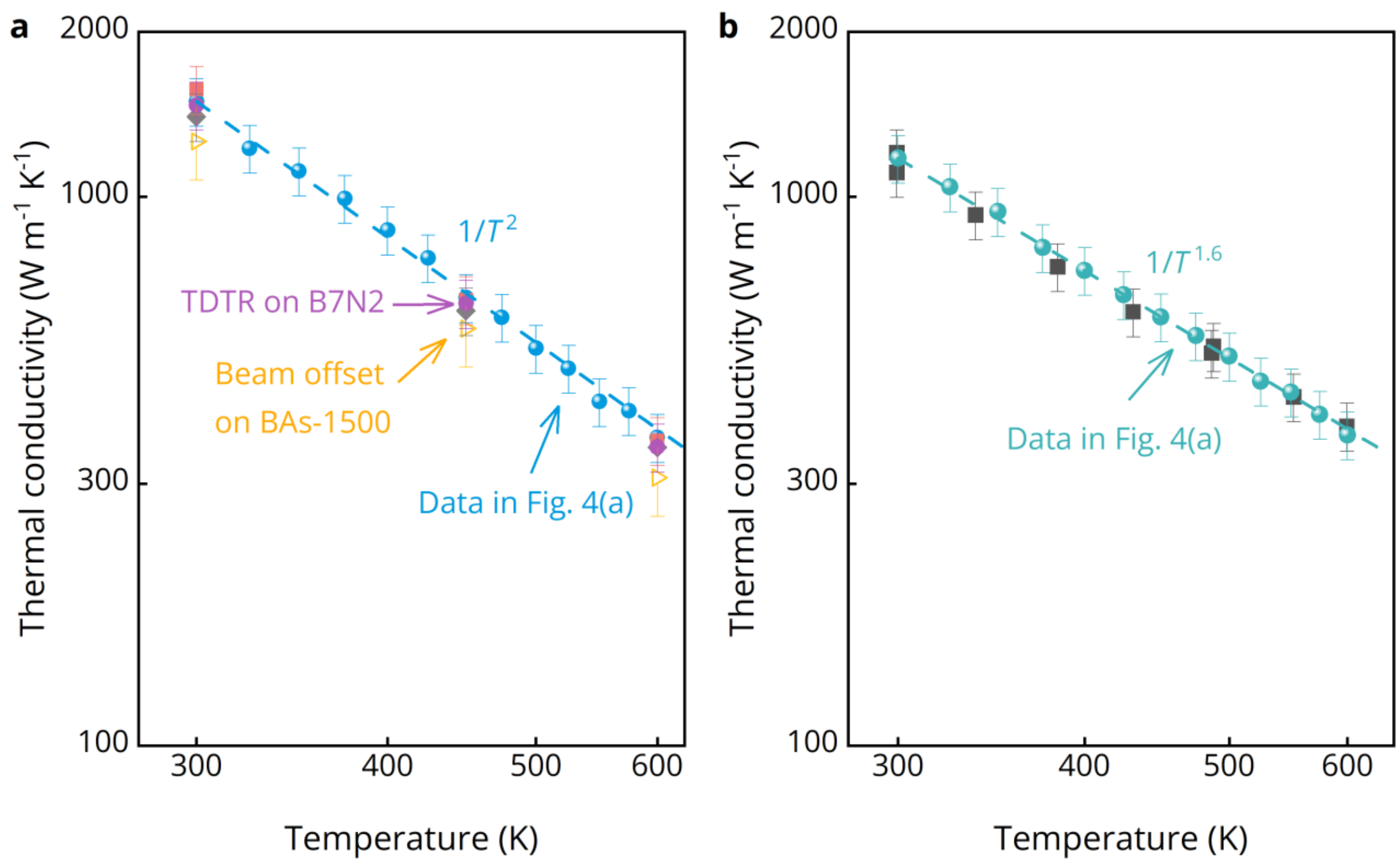

**Fig. 6.** Results of TDTR measurements performed to verify reproducibility of thermal conductivity results. (a) Summary of all the temperature dependent thermal conductivity measurements we performed on the BAs-1500 sample and sample B7N2. The blue circles in (a) correspond to the same data as the blue markers in Fig. 4(a). Gray and pink symbols are the results of additional measurements on the BAs-1500 sample measured on different dates. Yellow triangles are thermal conductivity of the same BAs-1500 sample measured using beam-offset TDTR measurements. Purple diamonds are the results of TDTR measurements on a different BAs crystal, labelled as B7N2, whose ambient thermal conductivity is also 1500 W m$^{-1}$ K$^{-1}$. (b) High-temperature TDTR measurements on two different BAs samples with $\Lambda_{300K}$ of 1200 W m$^{-1}$ K$^{-1}$. Dashed lines are power law fittings to the data. All the power law fittings follow the same procedure as described in Supplementary Materials[33].

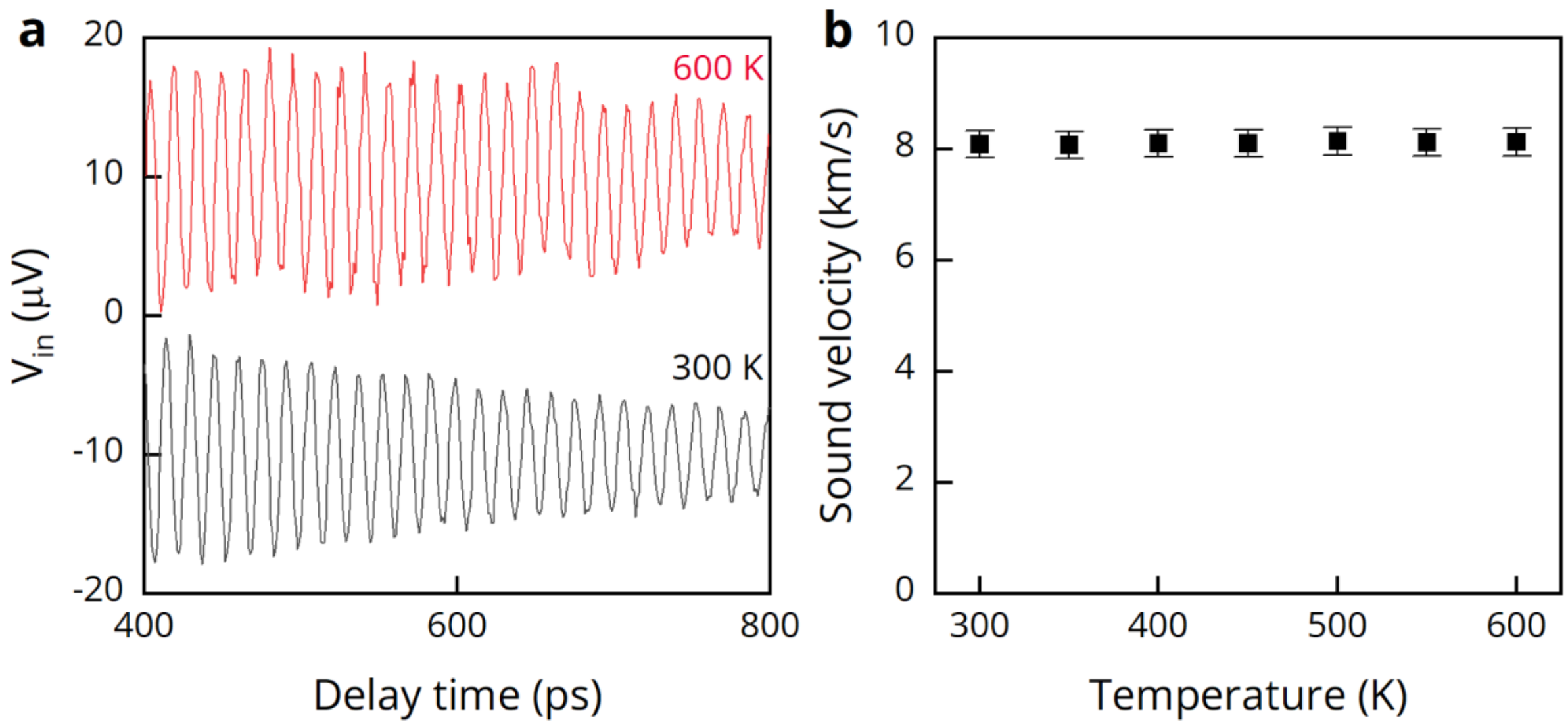

**Fig. 7.** Forced Brillouin scattering on BAs between 300 and 600 K. (a) Brillouin oscillation of BAs at 300 and 600 K. (b) The longitudinal sound velocity along [111] direction of BAs versus temperature derived from the Brillouin frequency. The observed Brillouin frequency is 65.2 GHz, consistent with prior reports of 65.2 GHz with 778 nm excitation wavelength[37,64]. The error bars are derived from the FWHM of the Brillouin spectrum (fast Fourier transform of panel a).

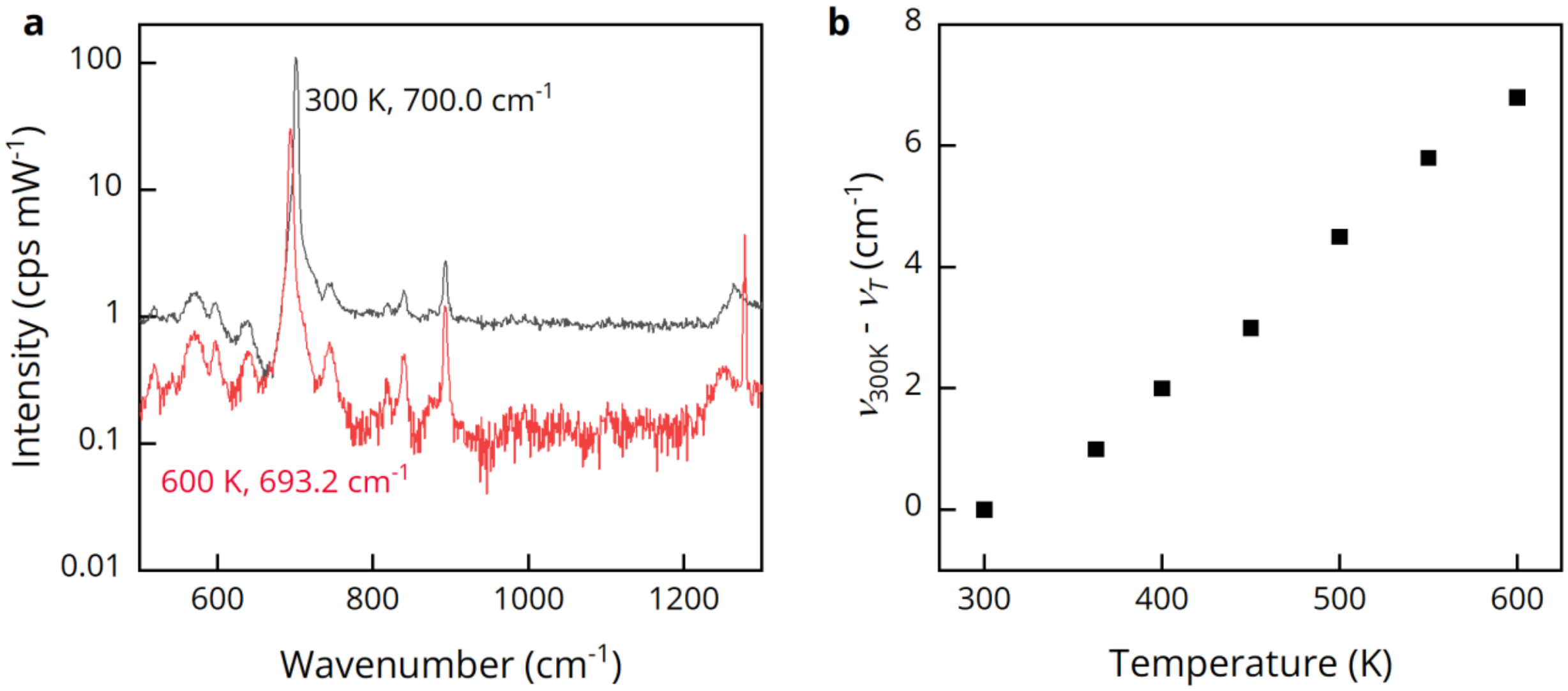

**Fig. 8**. Raman scattering on BAs between 300 and 600 K. (a) Raman spectra of BAs at 300 and 600 K. (b) Wavenumber change of the Raman peak relative to the 300 K value versus temperature.

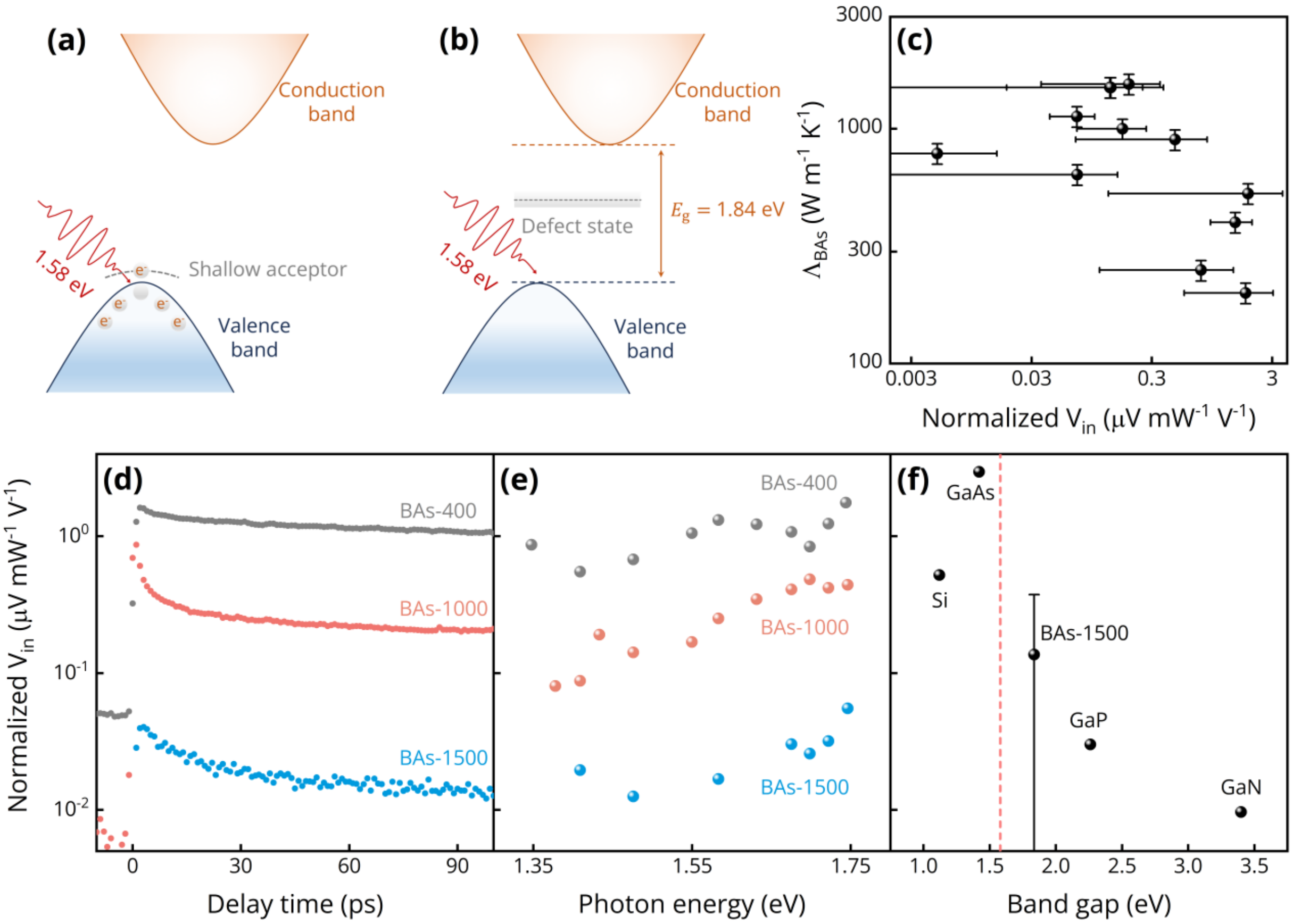

**Fig. 9.** Transient reflectivity microscopy measurements. Free carriers from shallow donors/acceptors, and deeper defect states in the band gap cause optical absorption at photon energies below the band gap. Optical absorption of the pump beam leads to measurable transient reflectivity microscopy signal. (a) and (b) are schematics showing the impact of a shallow acceptor and a deep defect state on absorption of energies below the band gap, respectively. (c) Thermal conductivity of BAs crystals versus TRM signals obtained from bare BAs samples with 1.58 eV incident photons. The TRM signal was collected at a fixed delay time of 90 ps. Normalized $V_{in}$ on the y-axis is the in-phase voltage measured by lock-in amplifier, normalized by the pump laser power and average voltage on the photodiode detector. (d) Time dependence of the transient reflectivity microscopy signal with 1.58 eV incident photons. (e) Transient reflectivity microscopy signals on three bare BAs samples as a function of excitation photon energy. (f) Transient reflectivity microscopy signals for bare Si, GaAs, BAs, GaP, and GaN single crystals. The vertical dashed line indicates laser energy.

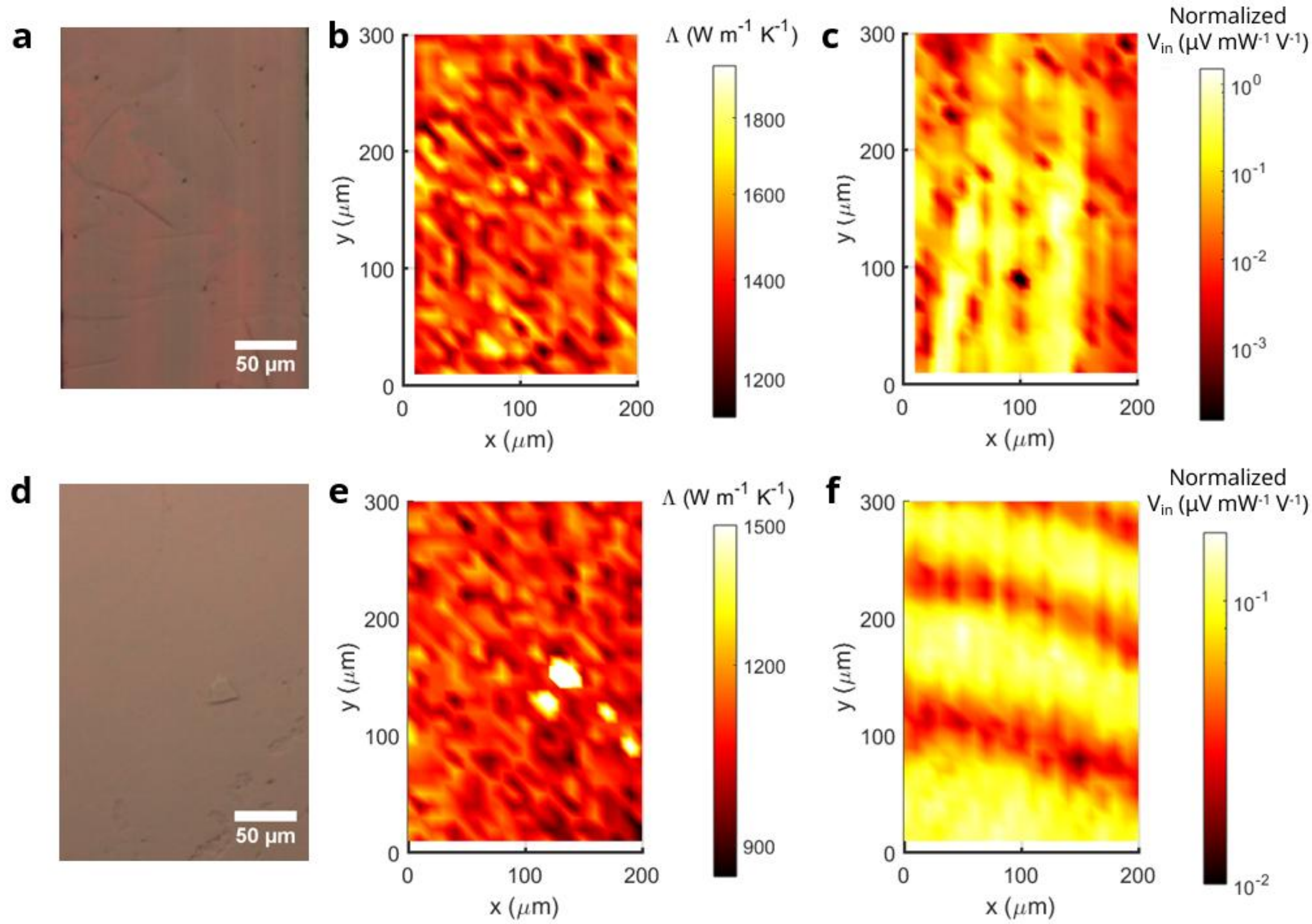

**Fig. 10.** Thermal conductivity and TRM maps on BAs crystals. (a) and (d) are optical images of two BAs samples whose ambient thermal conductivity are 1500 and 1000 W $m^{-1}$ $K^{-1}$, respectively. (b) and (e) are corresponding thermal conductivity maps. (c) and (f) are TRM maps on the two bare BAs samples correspondingly. (a), (b), and (c) are performed on the BAs-1500 sample.